\documentclass[12pt]{article}

\usepackage{amssymb}
\usepackage{amsmath}
\usepackage{bm}
\usepackage{upgreek}
\usepackage{graphicx}

\begin{document}
\begin{center}

{\bf CASIMIR THEORY OF THE RELATIVISTIC COMPOSITE STRING REVISITED, AND A FORMALLY RELATED PROBLEM IN SCALAR QFT}

\vspace{2cm}

  Iver Brevik\footnote{iver.h.brevik@ntnu.no}

\bigskip
Department of Energy and Process Engineering, Norwegian University of Science and Technology, N-7491 Trondheim, Norway

\bigskip

Revised version, June 2012

\begin{abstract}

The main part of this paper is to  present an updated review of
 the Casimir energy at zero and finite temperature for the transverse oscillations of a piecewise uniform closed string. We make use of three different  regularizations: the cutoff method, the complex contour integration method, and the zeta-function method. The string model is relativistic, in the sense that the velocity of sound is for each string piece set equal to the velocity of light. In this sense the
 theory is  analogous to the electromagnetic theory in a dielectric medium in which the product of permittivity and permeability is equal to unity (an isorefractive medium). We demonstrate how the formalism works for a two-piece string, and for a 2N-piece string, and show how in the latter case a compact recursion relation serves to facilitate the formalism considerably. The Casimir energy turns out to be negative, and the more so the larger  the number of pieces in the string.  The two-piece string is quantized in D-dimensional spacetime, in the limit when the ratio between the two tensions is very small. We calculate the free energy and other thermodynamic quantities, demonstrate scaling properties, and comment finally on the meaning of the Hagedorn critical temperature for  the two-piece string.Thereafter, as a novel development we present a scalar field theory for a real field in three-dimensional space in a potential rising linearly with a longitudinal coordinate $z$ in the interval $0<z<1$, and which is thereafter held constant on a horizontal plateau. The potential is taken as a rough model of the two-piece string potential under simplifying conditions, when the length ratio between the pieces is replaced formally with the mentioned length parameter $z$.\footnote{This article is dedicated to the 75th anniversary of Professor Stuart Dowker.}
\end{abstract}

\end{center}

\section{Introduction. The two-piece string}

The composite string model is a natural generalization of the conventional uniform string. Standard theory of closed strings - whatever the string is in Minkowski space or in superspace - assumes the string to be homogeneous along the length $\sigma$. The string tension $T$ is thus the same everywhere. The composite string model, taken in our context to be composed of two or more pieces of different materials, will be required to be relativistic in the sense that
 the velocity $v_s$ of transverse sound is everywhere  equal to the velocity of light,
\begin{equation}
v_s=\sqrt{T/\rho}=c=1. \label{1}
\end{equation}
Here $T$, as well as the mass density  $\rho$, refer to any of the  string pieces. At each junction there are two boundary conditions, namely (i) the transverse displacement $\psi=\psi(\sigma, \tau)$ is continuous, and (ii) the transverse force $T\partial \psi/\partial \sigma$ is continuous. The equation of motion is
\begin{equation}
\left(\frac{\partial^2}{\partial \sigma^2}-\frac{\partial^2}{\partial \tau^2}\right)\psi=0. \label{2}
\end{equation}
We consider first the two-piece string, composed of pieces whose lengths are $L_I$ and $L_{II}$.
With total string length $L=L_I+L_{II}$ we thus have
\begin{equation}
 \psi_I=\xi_Ie^{i\omega (\sigma-\tau)}+\eta_Ie^{-i\omega
(\sigma+\tau)}, \label{3}
\end{equation}
\begin{equation}
 \psi_{II}=\xi_{II}e^{i\omega
(\sigma-\tau)}+\eta_{II}e^{-i\omega(\sigma+\tau)}, \label{4}
\end{equation}
where the  continuity requirement
of $\psi$ at the junctions points yields
\begin{equation}
\xi_I+\eta_I=\xi_{II}e^{i\omega L}+\eta_{II}e^{-i\omega L}, \label{5}
\end{equation}
\begin{equation}
\xi_Ie^{i\omega L_I}+\eta_Ie^{-i\omega L_I}=\xi_{II}e^{i\omega
L_I}+\eta_{II}e^{-i\omega L_I}. \label{6}
\end{equation}
Continuity of transverse elastic force at the junctions yields
\begin{equation}
 T_I \frac{\partial \psi_I}{\partial \sigma}\Big|_{\sigma=0}=T_{II}\frac{\partial \psi_{II}}{\partial \sigma}\Big|_{\sigma=L}, \label{7}
 \end{equation}
 \begin{equation}
 T_I \frac{\partial \psi_I}{\partial \sigma}\Big|_{\sigma=L_I}
  =T_{II}\frac{\partial \psi_{II}}{\partial \sigma}\Big|_{\sigma=L_I}. \label{8}
  \end{equation}
  Now define
  \begin{equation}
   x=\frac{T_I}{T_{II}}, \quad s=\frac{L_{II}}{L_I}, \quad F(x)=\frac{4x}{(1-x)^2}. \label{9}
   \end{equation}
  The dispersion relation becomes
  \begin{equation}
 F(x)\sin^2\left( \frac{\omega L}{2}\right) +\sin \omega L_I\sin
\omega L_{II}=0, \label{10}
\end{equation}
The Casimir energy, describing the deviation from homogeneity,  can be written  formally as
\begin{equation}
E=E_{I+II}-E_{\rm uniform}=\frac{1}{2}\sum \omega_n-E_{\rm uniform}. \label{11}
\end{equation}
Since Eq.~(\ref{11}) is invariant under the substitution $x\rightarrow 1/x$, we can simply assume $x\leq 1$ in the following.
The two-piece string is sketched in Fig.~1.

\begin{figure}[htbp]
\makebox[\textwidth][c]{\includegraphics[width=5cm]{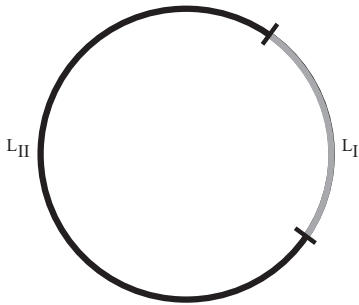}}
\begin{center}
\end{center}
\caption{The two-piece string, with piece lengths $L_I$ and $L_{II}$.}
\end{figure}
This model was introduced in 1990 \cite{brevik90}; cf. also the related paper \cite{li91}. The Casimir energy was calculated for various length ratios of the pieces.

From a physical point of view, there is a well- founded hope that this simple model can give insight in the properties of the vacuum state in two-dimensional quantum field theories in general. The issue of main interest for us here is, however, not to discuss possible practical applications of the model but rather to emphasize the great adaptivity the model has with respect to various regularization methods. The system is remarkably easy to regularize - this circumstance, in fact, supporting the expectation that the model may carry more physical information than has hitherto been recognized. In the paper \cite{brevik90} already mentioned, we made use of a  cutoff
regularization method whereby a function $f=\exp(-\alpha \omega)$
with $\alpha$ a small positive parameter was introduced. A second
regularization method is to make use of complex contour integrations.
 To our knowledge this
method was first applied to the composite string model by Brevik
and Elizalde \cite{brevik94}. A separate chapter is devoted to
this model in Elizalde's book on zeta functions \cite{elizalde95}.  One great
advantage of this method is that the {\it multiplicities} of the
zeros of the dispersion function are automatically taken care of. Third, one may regularize the system by using zeta function methods - as we know, Stuart Dowker has made significant contributions to zeta function theory and related topics in mathematical physics. For the specific piecewise uniform  string, it is the Hurwitz kind of zeta function that comes into play.

Instead of assuming only two pieces in the composite string, one can imagine that the string is composed of $2N$ pieces, all of the same length, such that the type I materials and the type II materials are alternating. Maintaining the same relativistic property as before, one will find  that also this kind of system is easily regularizable and tractable analytically in general.  There are by now several papers devoted to the study of the composite string in its various facets; cf. Refs.~\cite{brevik95,brevik96,bayin96,brevik97,berntsen97,brevik98,brevik99,brevik03,brevik11}.   As for possible physical applications of the model, we may also mention the paper of Lu and Huang \cite{lu98}, discussing the Casimir energy for a composite  Green-Schwarz superstring.

It ought to be stressed again that the regularizability of the string model relies upon the relativistic condition (\ref{1}). If this condition were abandoned, the formalism would at once be ambiguous and its predictions obscure. An interesting point in this context is to note the close relationship with the electromagnetic theory in a dielectric medium when the product of the permittivity and the permeability is equal to one (sometimes called an isorefractive medium). While Casimir theories for isorefractive media are easily tractable, the corresponding theory for an ordinary (non-isorefractive) medium would be difficult to construct. For recent papers along these lines, the reader may consult Refs.~\cite{brevik09,ellingsen09}.

 There have been further developments of the composite string theory in recent years. We may mention the extension to the so-called quantum spring model \cite{feng10}, where a helix boundary condition for a scalar massless field is imposed. The Casimir forces in the longitudinal and transverse directions for the spring are calculated, using zeta regularization, and the analogue to Hooke's law in elasticity is recovered when the pitch of the spring is small. Higher dimensions are also envisaged \cite{zhai11}. The fermionic Casimir effect with helix boundary condition is considered in Ref.~\cite{zhai11a}, and the scalar field with a  helix torus boundary condition in higher dimension is considered in Ref.~\cite{zhai11b}.

   As a proposal for future work, we mention that there may  be a connection between the phases of the piecewise uniform (super) string and the Bekenstein-Hawking entropy associated with this string. The entropy, as known, can be derived by counting black hole microstates, and it is natural to expect that the deviation from spatial homogeneity present in the composite string model could influence that sort of calculations.

   We ought here to mention the interesting analogy that seems to exist between the composite string model and the so-called quantum star graph model. Fulling {\it et al.} \cite{fulling07} recently studied vacuum energy and Casimir forces in one-dimensional quantum graphs (pistons), and found that the piston force could be attractive or repulsive depending on the number of edges. It may be that the mathematical similarities between these two kinds of theories reflect a deeper physical similarity also. This remains to be explored. Another more recent paper of Harrison and Kirsten \cite{harrison11} studies the zeta functions of quantum graphs.

 As mentioned above, the main part of this  paper is  an updated review of the main properties of the composite string model, at zero, and also at finite, temperature, making use of the different contour regularization methods. The convenience of the recursion formula in the $2N$-case, in particular, as treated in Sec.~6,  is in our opinion worth attention. The quantum theory of the two-piece string for the simplifying limiting case of very small tension ratio $x$ between the two pieces is highlighted, and the Hagedorn temperature is given for this kind of model.

 As a novel development, we give in Sect. 7 a scalar QFT for a real field $\phi$ residing in a potential $V(z)$ consisting of two linear pieces. The first piece, extending from $z=0$ to $z=1$ has a positive slope, while the second piece $z>1$ forms a horizontal plateau. The main idea behind this kind of potential comes from the two-piece string energy in the limiting case of extreme string tension, the string length ratio being replaced formally by a nondimensional length. The expression for the potential has to be simplified, however, in order to give real solutions for the field. We present the Green function for the problem, and give the expression for the field energy density as a function of $z$ for $z>1$.

 \section{Cutoff regularization}

 The simplest way to regularize the system is to introduce a convergence factor $f=e^{-\alpha \omega_n}, \quad \alpha \ll 1$ \cite{brevik90}, and to multiply the nonregularized expression for $E$ with $f$ before summing over the modes.

 We first consider a  uniform string, $(x=1)$, which implies
 \begin{equation}
 \omega_n=2\pi n/L, \quad n=1,2,3,... \label{12}
 \end{equation}
Since these modes are degenerate, we find for the zero-point energy
\begin{equation}
 E_{\rm uniform}=\frac{L}{2\pi \alpha^2}-\frac{\pi}{6L}+O(\alpha^2). \label{13}
 \end{equation}
The limiting case $x\rightarrow 0$ ($T_{II}$ assumed finite) leads to
two sequences of modes,
\begin{equation}
\omega_n=\pi n/L_I, \quad \omega_n=\pi n/L_{II}, \quad
n=1,2,3,...  \label{14}
\end{equation}
We then obtain
\begin{equation}
 E=-\frac{\pi}{24L}\left(s+\frac{1}{s}-2\right). \label{15}
 \end{equation}
If $s$ is an {\it odd} integer, the dispersion equation yields one degenerate branch, given by
\begin{equation}
 \omega L_I=\pi n, \label{16}
 \end{equation}
and $\frac{1}{2}(s-1)$ nondegenerate branches given by
\[ \omega L_I=\left\{ \begin{array}{ll}
\pi (n+\beta), \\
\pi (n+1-\beta),
\end{array}
\right. \]
with $n=1,2,3,..., ~0<\beta \leq 1/2$.

When $\alpha \rightarrow 0$:
\begin{equation}
 E=\frac{\pi s(s-1)}{12L}-\frac{\pi (s+1)}{4L}\sum_{i=1}^{(s-1)/2}[ \beta_i^2+(1-\beta_i)^2]; \label{17}
 \end{equation}
the cutoff term drops out.

If $s$ is {\it even}, we obtain analogously
\begin{equation}
 E=\frac{\pi s(2s+1)}{6L}-\frac{\pi (s+1)}{8L}\sum_{i=1}^s[\beta_i^2+(2-\beta_i)^2], \label{18}
 \end{equation}
where each $\beta_i$ lies in the interval $0<\beta_i \leq 1$.

 \section{Contour integration method}

This powerful method, in the context of Casimir calculations, dates back to van Kampen {\it et al.} \cite{vankampen68}. As mentioned, the method was first applied to the composite string system in Ref.~\cite{brevik94}. The starting point is the argument principle,
\begin{equation}
 \frac{1}{2\pi i}\oint \omega \frac{d}{d\omega}\ln
g(\omega)d\omega=\sum \omega_0-\sum \omega_\infty, \label{19}
\end{equation}
satisfied for any meromorphic function  $g(\omega)$, $\omega_0$ being the zeros and $\omega_\infty$ the poles of $g(\omega)$ inside the integration contour. The contour is taken to a semicircle of large radius $R$ in the right half $\omega$ plane, closed by a straight line from $\omega=iR$ to $\omega=-iR$. An advance of the method, as mentioned, is that the multiplicities are taken care of automatically.

It is convenient to choose
\begin{equation}
g(\omega)=\frac{F(x)\sin^2[(s+1)\omega L_I/2]+\sin (\omega L_I)
\sin(s\omega L_I)}{F(x)+1},  \quad {\rm with}~ s \geq 1. \label{20}
\end{equation}
This choice allows us to perform partial integrations in the energy integral without encountering any divergences in the boundary terms when $R\rightarrow \infty$. The final result, with $\omega =i\xi$, becomes
\begin{equation}
E=\frac{1}{2\pi}\int_0^\infty \ln \left| \frac{F(x)+\frac{\sinh (\xi L_I)\sinh (s\xi L_I)}{\sinh^2[(s+1)\xi L_I/2]}}{F(x)+1} \right| d\xi, \label{21}
\end{equation}
This expression holds for all values of $s$, not necessarily integers. As the substitution $s\rightarrow 1/s$ leaves the expression invariant, we may restrict ourselves to the interval $s\geq 1$.  If the tension ratio $x\rightarrow 0$, we recover the expression (\ref{15}).

At finite temperature
 $T$, where  $\xi_n=2\pi nT$ with $n=0,1,2,3,...$, we get the corresponding expression $(k_B=1)$
 \begin{equation}
 E(T)=T{\sum_{n=0}^\infty}'\ln
\Big|
\frac{F(x)+\frac{\sinh(\xi_nL_I)\sinh(s\xi_nL_I)}{\sinh^2[(s+1)\xi_nL_I/2]}}{F(x)+1}
\Big|, \label{22}
\end{equation}
where the prime means that the mode  $n=0$ is counted with half weight.

 We may define two characteristic frequencies in the problem: (i) the thermal
 frequency
 $ \omega_T=T = \xi_1/(2\pi)$, and (ii) the geometric frequency
$ \omega_{\rm geom}=2\pi/L_I$. The case of high temperatures corresponds to
$ \omega_T/\omega_{\rm geom} \geq 1$, when we can approximate
\begin{equation}
 E(T)=\frac{1}{2}T\ln \Big| \frac{F(x)+4s/(s+1)^2}{F(x)+1} \Big|. \label{23}
 \end{equation}
Thus if  "our" universe (I) is small and the "mirror" universe (II) large ($s\rightarrow \infty$), we have
\begin{equation}
 E(T)=-\frac{1}{2}T\ln \left|1+F(x)^{-1}\right|. \label{24}
 \end{equation}
In the case of low temperatures,
$ \omega_T/\omega_{\rm geom} \ll 1$, a large
 number of Matsubara frequencies $\xi_n$ is necessary.

 \section{Zeta-function method}

General treatises on this elegant regularization method can be found in Refs.~\cite{elizalde95} and \cite{elizalde94}. The first application to the composite string was made by Li {\it et al.} \cite{li91}. The zeta function of most use in this case is not the Riemann function $\zeta(s)$ but instead the Hurwitz function $\zeta_H(s,a)$, originally defined as
\begin{equation}
 \zeta_H(s,a)=\sum_{n=0}^\infty (n+a)^{-s}, \quad 0<a<1, \quad
 \Re s>1. \label{25}
 \end{equation}
 For practical purposes one  needs the property
 \begin{equation}
  \zeta_H(-1,a)=-\frac{1}{2}(a^2-a+\frac{1}{6}) \label{26}
  \end{equation}
 of the analytically continued function. The Hurwitz function in the form (\ref{25}) is defined for $\Re s>1$; it is a meromorphic function with a simple pole in $s=1$. If $\Re s$ is different from unity, the Hurwitz function can be analytically continued to the whole complex plane. In our case, Eq.~({26}) gives the finite value of the Hurwitz function at $s=-1$.

 When using the zeta-function method one has to determine the eigenvalue spectrum explicitly. This is the same property as one encounters when using the cutoff method.

 Consider first the uniform string: in this case the Riemann function $\zeta(s)$ is sufficient, giving the zero-point energy
 \begin{equation}
 E_{\rm uniform}=\frac{2\pi}{L}\zeta_R(-1)=-\frac{\pi}{6L}, \label{27}
 \end{equation}
 in agreement with the finite part of Eq.~(\ref{13}). Consider next the composite string, assuming $s$ to be an odd integer. Inserting the degenerate branch eigenvalue spectrum (\ref{16}) we get
 \begin{equation}
 E_{\rm degenerate~branch}=-\frac{\pi}{12L_I}. \label{28}
 \end{equation}
 Using the corresponding forms for the double branches we obtain analogously
 \[ E_{\rm double~branch} =\frac{\pi}{2L_I}[ \zeta_H(-1,\beta)+\zeta_H(-1,1-\beta)] \]
 \begin{equation}
 =\frac{\pi}{6L_I}-\frac{\pi}{4L_I}[\beta^2+(1-\beta)^2]. \label{29}
 \end{equation}
 Summing (\ref{29}) over the $\frac{1}{2}(s-1)$ double branches, and adding (\ref{28}), we obtain for the composite string's zero-point energy
 \begin{equation}
  E_{I+II}=\frac{\pi(s-2)}{12L_I}-\frac{\pi}{4L_I}\sum_{i=1}^{(s-1)/2}[\beta_i^2+(1-\beta_i)^2]. \label{30}
  \end{equation}
Now subtracting off the expression (\ref{27}) we get
\begin{equation}
  E=E_{I+II}-E_{\rm uniform}=\frac{\pi s(s-1)}{12L}-\frac{\pi (s+1)}{4L} \sum_{i=1}^{(s-1)/2}[\beta_i^2+(1-\beta_i)^2], \label{31}
  \end{equation}
  in agreement with Eq.~(\ref{17}).

  The case of even integers $s$ is treated analogously. The zeta-function method is somewhat easier to implement than the cutoff method.

\section{  Oscillations of the two-piece string in  $D$-dimensional spacetime. Quantization}

 Our aim is now to  sketch the essentials of the quantum theory, for  the two-piece string. To allow for a correspondence to the superstring, we allow the number of flat spacetime dimensions $D$ to be an arbitrary integer. In accordance with usual practice, we put now  $L=L_I+L_{II}=\pi$. The theory will be based on two simplifying assumptions:

(i)  The string tension ratio $x\rightarrow 0$. The dispersion relation (\ref{10}) leads in this case to two different branches of solutions, namely the first branch obeying
\begin{equation}
\omega_n(s)=(1+s)n, \label{43}
\end{equation}
and the second branch obeying
\begin{equation}
\omega_n(s^{-1})=(1+s^{-1})n, \label{44}
\end{equation}
with $n=\pm 1, \pm 2, \pm 3,....$

(ii) The second assumption is that the length ratio $s$ is an integer, $s=1,2,3,...$.

\bigskip

Let now $ X^\mu(\sigma, \tau)$ with $\mu=0,1,2,..(D-1)$ be the coordinates on the world sheet. For each branch
\begin{equation}
 X^\mu=x^\mu+\frac{p^\mu\tau}{\pi \bar{T}(s)}+\theta(L_I-\sigma)X_I^\mu+\theta(\sigma-L_I)X_{II}^\mu, \label{45}
 \end{equation}
 where $\theta(x)$ is the step function, $x^\mu$ the center of mass position, and $p^\mu$ the total  momentum of the string. The mean tension in the actual limit is $ \bar{T}(s)={T_{II}s/(1+s)}$ (we assume $T_{II}$ finite). The string's translational energy is
$ p^0=\pi \bar{T}(s)$. In the following we consider the first branch only.

In region I we make the expansion
\begin{equation}
 X_I^\mu=\frac{i}{2\sqrt{\pi T_I}}\sum_{n\neq
0}\frac{1}{n}\left[\alpha_n^\mu(s)e^{i(1+s)n(\sigma-\tau)}+\tilde{\alpha}_n^\mu(s)e^{-i(1+s)n(\sigma+\tau)}\right],
 \label{46}
 \end{equation}
 where $ \alpha_{-n}^\mu=(\alpha_n^\mu)^*, \,
\tilde{\alpha}_{-n}^\mu=(\tilde{\alpha}_n^\mu)^*$. The action can be expressed as
\begin{equation}
 S=-\frac{1}{2}\int d\tau d\sigma
T(\sigma)\eta^{\alpha\beta}\partial_\alpha X^\mu\partial_\beta
X_\mu, \label{47}
\end{equation}
 where $ T(\sigma)=T_I+(T_{II}-T_I)\theta(\sigma-L_I)$. As the conjugate momentum is $P^\mu(\sigma)=T(\sigma)\dot{X}^\mu,$ we obtain the Hamiltonian
\begin{equation}
H=\int_0^\pi [P_\mu(\sigma)\dot{X}^\mu-L]d\sigma=\frac{1}{2}\int_0^\pi T(\sigma)(\dot{X}^2+{X'}^2)d\sigma.
\label{48}
\end{equation}
The fundamental condition is that  $H=0$ when applied to physical states.

The corresponding expansion of the first branch in
region II is
\begin{equation}
 X_{II}^\mu=\frac{i}{2\sqrt{\pi T_I}}\sum_{n\neq 0}
\frac{1}{n}\gamma_n^\mu(s)e^{-i(1+s)n\tau}\cos [(1+s)n\sigma], \label{49}
\end{equation}
with
\begin{equation}
 \gamma_n^\mu(s)=\alpha_n^\mu(s)+\tilde{\alpha}_n^\mu(s), \quad
n\neq 0. \label{50}
\end{equation}
The condition $x\rightarrow 0 $ means that there are only  standing waves
in region II.

We may now introduce light-cone coordinates $ \sigma^-=\tau-\sigma, \, \sigma^+=\tau+\sigma$. Some calculation shows that the total Hamiltonian can be written as a sum of two parts,
\begin{equation}
 H=H_I+H_{II}, \label{51}
 \end{equation}
where
\begin{equation}
 H_I=\frac{1+s}{4}\sum_{-\infty}^\infty [\alpha_{-n}(s)\cdot
\alpha_n(s)+\tilde{\alpha}_{-n}(s)\cdot \tilde{\alpha}_n(s)], \label{52}
\end{equation}
\begin{equation}
 H_{II}=\frac{s(1+s)}{8x}\sum_{-\infty}^\infty
\gamma_{-n}(s)\cdot \gamma_n(s). \label{53}
\end{equation}
The mass $M$ of the string determined by $M^2=-p^\mu p_\mu$,
\begin{equation}
 M^2=\pi T_{II}s\sum_{n=1}^\infty \left[ \alpha_{-n}(s)\cdot
\alpha_n(s)+\tilde{\alpha}_{-n}(s)\cdot
\tilde{\alpha}_n(s)+\frac{s}{2x}\gamma_{-n}(s)\cdot
\gamma_n(s)\right]. \label{54}
\end{equation}
Recall that this is the contribution from first branch only. The expression is valid
for even/odd values of $s$.

Consider now the quantization of the first branch modes. We impose
the conditions
\begin{equation}
 T_I[ \dot{X}^\mu(\sigma,
\tau),X^\nu(\sigma',\tau)]=-i\delta(\sigma-\sigma')\eta^{\mu\nu}
\label{55}
\end{equation}
in region I, and
\begin{equation}
 T_{II}[ \dot{X}^\mu(\sigma,
\tau),X^\nu(\sigma',\tau)]=-i\delta(\sigma-\sigma')\eta^{\mu\nu}
\label{56}
\end{equation}
in region II (the other commutators vanish). Then introducing
creation and annihilation operators via
\begin{equation}
\alpha_n^\mu(s)=\sqrt{n}a_n^\mu(s), \quad
\alpha_{-n}^\mu(s)=\sqrt{n}{a_n^\mu}^ \dagger (s), \label{57}
\end{equation}
\begin{equation}
 \gamma_n^\mu(s)=\sqrt{4nx}c_n^\mu(s), \quad
\gamma_{-n}(s)=\sqrt{4nx}{c_n^\mu}^\dagger (s), \label{58}
\end{equation}
one arrives at the conventional commutation relations
\begin{equation}
 [a_n^\mu(s), {a_m^\nu}^\dagger (s)]=\delta_{nm}\eta^{\mu\nu},
 \quad
 [c_n^\mu(s),{c_m^\nu}^\dagger (s)]=\delta_{nm}\eta^{\mu\nu},
 \label{59}
 \end{equation}
for $n,m \geq 1$.

Now introduce $t(s)$ as
\begin{equation}
t(s)=\pi \bar{T}(s)=\frac{\pi s}{1+s}T_{II}, \label{60}
\end{equation}
and put $D=26$, the usual dimension for the bosonic string. The
condition  $H=H_I+H_{II}=0 $ leads to
\[
 M^2=t(s)\sum_{i=1}^{24}\sum_{n=1}^\infty
\omega_n(s)[{a_{ni}}^\dagger (s)a_{ni}(s)+{\tilde{a}_{ni}}^\dagger
(s)\tilde{a}_{ni}(s)-2] \]
\begin{equation}
+2st(s)\sum_{i=1}^{24}\sum_{n=1}^\infty \omega_n(s)[
{c_{ni}}^\dagger (s)c_{ni}(s)-1], \label{61}
\end{equation}
and the free energy becomes
\[ F=-\frac{1}{24}(s+\frac{1}{s}-2)-2^{-40}\pi^{-26}t(s)^{-13}\int_0^\infty \frac{d\tau_2}{\tau_2^{14}}\int_{-1/2}^{1/2}d\tau_1 \]
\begin{equation}
 \times \left[ \theta_3\left(0\Big|
\frac{i\beta^2t(s)}{8\pi^2\tau_2}\right)-1\right]\Big|
\eta[(1+s)\tau]\Big|^{-48}\eta[s(1+s)(\tau-\bar{\tau})]^{-24}
\label{62}
\end{equation}
Here $ \tau=\tau_1+i\tau_2$ is the
 $\rm Teichm\ddot{u}ller$ parameter,
 \begin{equation}
 \eta(\tau)=e^{\pi i\tau/12}\prod_{n=1}^\infty \left( 1-e^{2\pi
in\tau}\right)  \label{63}
\end{equation}
is the Dedekind $\eta$-function, and
\begin{equation}
 \theta_3(v|x)=\sum_{n=-\infty}^\infty e^{ixn^2+2\pi i
 vn}\label{64}
 \end{equation}
 is the Jacobi $\theta_3$ - function. From this the thermodynamic
 quantities such as internal energy $U$ and entropy $S$ can be
 calculated,
 \begin{equation}
 U=\frac{\partial(\beta F)}{\partial \beta}, \quad
S=\beta^2\frac{\partial F}{\partial \beta}. \label{65}
\end{equation}
Finally, let us consider the limiting case in which one of the string pieces is much shorter than the other. Physically this case is of interest, since it corresponds to a point mass sitting on a string. Since we have assumed that $s\geq 1$, this case corresponds to $s\rightarrow \infty$. We let the tension $T_{II}$ be fixed, though arbitrary. It is seen that the critical temperature goes to infinity so that the free energy $F$ is always ultraviolet finite. In this limit we obtain \cite{brevik98}
\[ F_{(\beta \rightarrow 0)} =-\frac{s}{24}-(8\pi^3T_{II})^{-13}\int_0^\infty \frac{d\tau_2}{\tau_2^{14}}\int_{-1/2}^{1/2}d\tau_1 \]
\begin{equation}
\times \exp \left( -\frac{\beta^2 T_{II}}{8\pi \tau_2}\right) |\eta [(1+s)\tau]|^{-48} \,\eta[s(1+s)(\tau-\bar{\tau})]^{-24}. \label{X}
\end{equation}
The second term goes to zero for large $s$. Physically speaking, the linear dependence of the first term  reflects that the Casimir energy of a little piece of string embedded in an essentially infinite string is for dimensional reasons inversely proportional to the length $L_I$. Thus
\begin{equation}
F \propto 1/L_I =(1+s)/\pi \approx s/\pi. \label{67}
\end{equation}

 \section{2N-piece string}

 In the same way one can consider the Casimir theory for a string of length $L$ divided into three pieces, all of the same length. The theory for this case was given in Refs.~\cite{brevik96} and \cite{berntsen97}. Here, we shall consider instead a string divided into $2N $ pieces of equal length, of alternating type I (type II) material. The string is relativistic, i.e. it satisfies the condition (\ref{1}), as above. The basic formalism for arbitrary integers $N$ was established in Ref.~\cite{brevik95}, but the Casimir energy was there calculated in full only for the case $N=2$. A full calculation was given in Ref.~\cite{brevik97}, cf. also Ref.~\cite{berntsen97}. A key point in Ref.~\cite{brevik97} was the derivation of a new recursion formula, applicable for general integers $N$.

 Now define the symbols $p_N$ and $\alpha$,
 \begin{equation}
 p_N=\frac{\omega L}{N}, \quad \alpha=\frac{1-x}{1+x},  \label{32}
 \end{equation}
 and recall that $ x=T_I/T_{II}$.
The eigenfrequencies are determined from
\begin{equation}
 \rm{Det}\left[ {\bf M}_{2N}(x,p_N)-{\bf 1} \right]=0, \label{33}
 \end{equation}
 where
 \[
 {\bf M}_{2N}(x,p_N)=\left[ \frac{(1+x)^2}{4x}\right]^N {\bf \Lambda}^N(\alpha, p_N), \]
 \begin{equation}
 {\bf \Lambda}(\alpha, p)=\left( \begin{array}{ll}
a & b \\
b^* & a^*
\end{array}
\right), \label{34}
\end{equation}
and
\[ a=e^{-ip}-\alpha^2, \quad b=\alpha (e^{-ip}-1). \]
The expression (\ref{34}) relies upon a compact recursion formula which serves to simplify the calculation (more details are given in Ref.~\cite{brevik97}). One can now calculate the eigenvalues of $\bf \Lambda$, and express the elements of ${\bf M}_{2N}$ as powers of these.

The Casimir energy can now be found. By means of  contour integration we can write, for arbitray $x$ and arbitray integers $N$, at zero temperature,
\begin{equation}
 E_N(x)=\frac{N}{2\pi L}\int_0^\infty \ln \left| \frac{2(1-\alpha^2)^N-[\lambda_+^N(iq)+\lambda_-^N(iq)]}{4\sinh^2(Nq/2)} \right| dq, \label{35}
 \end{equation}
where $\lambda_\pm$ are eigenvalues of $\bf \Lambda$, for imaginary arguments $iq$, of the dispersion equation. Explicitly,
\begin{equation}
 \lambda_\pm(iq)=\cosh q-\alpha^2 \pm [(\cosh q-\alpha^2)^2-(1-\alpha^2)^2]^{1/2}. \label{36}
 \end{equation}
Evaluation of the integral shows that  $E_N(x)<0,~|E_N(x)|$ increasing with increasing $N$. A string is thus in principle able to diminish its zero-point energy by dividing itself into a larger number of pieces of alternating type I (or II) material. If this property has physical significance, is at present unknown.

If $x\rightarrow 0$ the integral can be solved exactly,
\begin{equation}
 E_N(0)=-\frac{\pi}{6L}(N^2-1). \label{37}
 \end{equation}
 The generalization of the expression (\ref{35}) to the case of finite temperatures is easily achieved following the same procedure as above.

 As  an alternative method one can instead of contour integration make use of the zeta-function method. One then has to determine the spectrum explicitly and thereafter put in the degeneracies by hand. The latter method is therefore most suitable for low $N$.

 \bigskip

{\bf Scaling invariance}

\vspace{0.5cm}
A rather unexpected scaling invariance property of the Casimir energy becomes apparent if we examine the behavior of the function $f_N(x)$ defined by
\begin{equation}
 f_N(x)=\frac{E_N(x)}{E_N(0)}, \quad 0<f_N(x)<1. \label{38}
 \end{equation}
 This function lies between zero and one. By calculating $E_N(x)$ (usually numerically) as a function of $x$ for a fixed value of $N$, we find that the resulting curve for $f_N(x)$ is practically the {\it same}, irrespective of the value of $N$, as long as $N\geq 2$. (The case $N=1$ is exceptional, since $E_1(x)=0$.) Numerical trials show that the analytic form
 \begin{equation}
  f_N(x) \rightarrow f(x)=(1-\sqrt{x})^{5/2}  \label{39}
  \end{equation}
  is a useful approximation, especially in the interval $0<x<0.45$ \cite{brevik99}.

At finite temperature the expression for the Casimir energy
becomes
\begin{equation}
E_N^T(x)=T{\sum_{n=0}^\infty}' \ln \left|
\frac{2(1-\alpha^2)^N-[\lambda_+^N(i\xi_n L/N)+\lambda_-^N(i\xi_n
L/N)]}{4\sinh^2(\xi_nL/2)}\right|, \label{40}
\end{equation}
where $\lambda_\pm (i\xi_nL/N)$ are given by Eq.~(\ref{36}) with
$q\rightarrow q_n=\xi_nL/N$. It is here useful to note that
\begin{equation}
\lambda_+(iq_n)+\lambda_-(iq_n)=2(\cosh q_n-\alpha^2).
\label{41}
\end{equation}
There are several special cases of interest. First, if the string
is uniform ($x=1$), we get $E_N^T(1)=0$. This is as expected, as
the Casimir energy is a measure of the string's inhomogeneity. If
$N=1$, $x$ arbitrary, we also get a vanishing result,
$E_1^T(x)=0$. In particular, if $x\rightarrow 0$ we get the simple
formula
\begin{equation}
E_N(0)=2T{\sum_{n=0}^\infty }' \ln
\left|\frac{2^N\sinh^N(\xi_nL/2N)}{2\sinh(\xi_nL/2)}\right|.
\label{42}
\end{equation},

\section{A formally related problem in scalar quantum field theory}

Consider the following problem in the quantum theory of a massless scalar field $\phi(x)$ ($x={\bf r}, t$) in three-dimensional space if there is a potential $V=V(z)$ varying in the $z$ direction only,
\begin{equation}
V(z)=-\left( z+\frac{1}{z}-2\right), \quad 0\leq z <\infty. \label{A}
\end{equation}
One may ask: is it possible to calculate the field energy density $u(z)$ analytically as a function of $z$ for such a case? We see that the potential $V$ reflects actually our previous expression (\ref{15}) for the Casimir energy $E$ of a two-piece string, in the limit when the string tension ratio $x=T_I/T_{II}$ goes to zero, if the length ratio $s=L_{II}/L_I$ is replaced with the "length" $z$. The ansatz (\ref{A}) is given in a nondimensional form, for simplicity. The zero value of $V$ at $z=1$ corresponds to the previous zero Casimir energy $E$ at $s=1$.

Evidently, the relationship with our Casimir theory is quite formal. Nevertheless, we find it of interest to explore the QFT problem based upon Eq.~(\ref{A}) in its own right , so also because of the recent interest in this kind of scalar field models in the recent literature.

Let us first look at some consequences of using the potential (\ref{A}) as it stands. The field equation
\begin{equation}
\left(\Box +z+\frac{1}{z}-2\right)\phi(x)=0, \label{B}
\end{equation}
when combined with the Fourier transform
\begin{equation}
\phi(x)=\int \frac{d\omega}{2\pi}\frac{d\bf k}{(2\pi)^2}e^{-i\omega t+i{\bf k\cdot r}} \phi(z), \label{C}
\end{equation}
leads to the following equation
\begin{equation}
\phi''(z)+\left( z+\frac{1}{z}-\kappa^2-2\right)\phi(z)=0. \label{D}
\end{equation}
Here $\bf k$ means the wave vector in the $xy$ plane, transverse to the $z$ axis, and $\kappa^2$ is defined as
\begin{equation}
\kappa^2={\bf k}^2-\omega^2.\label{E}
\end{equation}
We have looked for exact solutions of Eq.~(\ref{D}) using Mathematica, without finding an exact solution. It is of interest nevertheless to consider some limiting cases. First, if $z\rightarrow 0$ the field equation reduces to
\begin{equation}
\phi''(z)+\frac{1}{z}\phi(z)=0, \label{F}
\end{equation}
admitting the solution
\begin{equation}
\phi(z) \propto \sqrt{z}J_1(2\sqrt z),\label{G}
\end{equation}
(infinities at the origin  discarded). Here $J_1$ is the Bessel function of the first kind. Next, in the region where $z$ lies around unity we get
\begin{equation}
\phi''(z)-\kappa^2\phi(z)=0, \label{H}
\end{equation}
 implying that $\phi$ contains linear combinations of $e^{\kappa z}$ and $e^{-\kappa z}$. Finally, as $z\rightarrow \infty$ the reduced equation takes the form
\begin{equation}
\phi''(z)+z\phi(z)=0, \label{I}
\end{equation}
which has complex solutions, $Ai[(-1)^{1/3}z]$ and $Bi[(-1)^{1/3}z]$, $Ai$ and $Bi$ being the Airy functions. The complex nature of these solutions are related to the negative slope of the potential $V$ when $z>1$. In our case such solutions are not of interest; we are looking for real solutions for $\phi(z)$.

We thus conclude that in order to construct a meaningful QFT for the real field, we have to replace the expression for $V$ above with a simpler form. Our choice in the following will be the most simple and natural one, namely to choose a linear wall, increasing from $z=0$ with a positive constant slope to a maximum value of $V=0$ at $z=1$. For higher $z$ we assume there to be a constant plateau. Thus, our potential assumed in the following will have the form
\begin{equation*}
 V(z)=z-1, \quad 0\leq z<1,
 \end{equation*}
\begin{equation}
V(z)=0, \quad z>1. \label{J}
\end{equation}
In the first equation the slope is chosen equal to unity for simplifying reasons. Our choice (\ref{J}) is of essentially the same form as that considered recently by Milton \cite{milton11}; cf. also related papers by Bouas {\it et al.} \cite{bouas11}.

The Green function $G(x,x')$ with $V(z)$  given by Eq.~(\ref{J}) satisfies the governing equation
\begin{equation}
\left( \frac{\partial^2}{\partial t^2}-\nabla^2+V(z)\right)G(x,x')=\delta(x-x'). \label{K}
\end{equation}
With the Fourier transform
\begin{equation}
G(x,x')=\int \frac{d\omega}{2\pi}\frac{d\bf k}{(2\pi)^2}e^{-i\omega (t-t')+i \bf k\cdot (r-r')}g(z,z'), \label{L}
\end{equation}
we get the following equation for the Fourier component
\begin{equation}
\left( \frac{\partial^2}{\partial z^2}-\kappa^2-V(z)\right) g(z,z')=-\delta (z-z'). \label{M}
\end{equation}
We perform a complex frequency rotation $\omega \rightarrow i\zeta$, so that $\kappa^2 \rightarrow {\bf k}^2+\zeta^2$.

Assume in the following that $z'$ lies on the horizontal plateau, $z'>1$. Then,
\begin{equation*}
 0\leq z<1: \quad g(z,z')=A(z'){\rm Ai}(\kappa^2-1+z)+B(z'){\rm Bi}(\kappa^2-1+z),
 \end{equation*}
\begin{equation}
\noindent z>1: \quad g(z,z')=\frac{1}{2\kappa}e^{-\kappa |z-z'|}+C(z')e^{-\kappa(z-1)}, \label{N}
\end{equation}
where we have adopted the boundary condition $G\rightarrow 0$ at $z\rightarrow \infty$. At $z=0$ we impose the Dirichlet condition:
\begin{equation}
A(z'){\rm Ai}(\kappa^2-1) +B(z'){\rm Bi} (\kappa^2-1)=0. \label{O}
\end{equation}
To determine the functions $A, B$ and $C$ we need two more boundary conditions, namely that $G$ itself as well as its derivative are continuous at $z=1$:
\begin{equation*}
 A(z'){\rm Ai}(\kappa^2)+B(z'){\rm Bi}(\kappa^2)=\frac{1}{2\kappa}e^{-\kappa z'}+C(z'),
 \end{equation*}
\begin{equation}
A(z'){\rm Ai}'(\kappa^2)+B(z'){\rm Bi}'(\kappa^2)=\frac{1}{2}e^{-\kappa z'}-\kappa C(z'). \label{P}
\end{equation}
The solution to this set of equations can be written as
\begin{equation*}
A(z')=\frac{{\rm Bi}(\kappa^2-1)}{{\rm Ai}(\kappa^2){\rm Bi}(\kappa^2-1)-{\rm Ai}(\kappa^2-1){\rm Bi}(\kappa^2)} \,\frac{e^{-\kappa z'}}{Q},
\end{equation*}
\begin{equation*}
B(z')= -\frac{{\rm Ai}(\kappa^2-1)}{{\rm Ai}(\kappa^2){\rm Bi}(\kappa^2-1)-{\rm Ai}(\kappa^2-1){\rm Bi}(\kappa^2)} \,\frac{e^{-\kappa z'}}{Q},
\end{equation*}
\begin{equation}
C(z')=-\left(\frac{1}{2}-\frac{1}{Q}\right) e^{-\kappa z'}, \label{Q}
\end{equation}
where we have defined $Q$ as
\begin{equation}
Q=\kappa +\frac{{\rm Ai}'(\kappa^2){\rm Bi}(\kappa^2-1)-{\rm Ai}(\kappa^2-1){\rm Bi}'(\kappa^2)}{{\rm Ai}(\kappa^2){\rm Bi}(\kappa^2-1)-{\rm Ai}(\kappa^2-1){\rm Bi}(\kappa^2)}. \label{R}
\end{equation}
These expressions are complicated. It is of interest to consider approximate values in limiting cases. Let us assume the case of high frequencies, implying that $\kappa^2={\bf k}^2+\zeta^2 \gg 1$. Then, we have as rough approximations \cite{NIST}
\begin{equation}
{\rm Ai}(x) \sim \frac{e^{-\xi}}{2\sqrt{\pi}x^{1/4}}, \quad {\rm Bi}(x)\sim \frac{e^\xi}{\sqrt{\pi}x^{1/4}}, \label{S}
\end{equation}
where $\xi=\frac{2}{3}x^{3/2}$. Together with the Wronskian for general argument, $W\{ {\rm Ai}(x), {\rm Bi}(x)\}=\frac{1}{\pi}$, this yields
\begin{equation*}
A(z') \sim -\sqrt{\frac{\pi}{\kappa}}\,e^{-\kappa (z'+1)}\exp \left[\frac{2}{3}(\kappa^2-1)^{3/2}\right],
\end{equation*}
\begin{equation*}
B(z')\sim \frac{1}{2}\sqrt{\frac{\pi}{\kappa}}\,e^{-\kappa (z'+1)}\exp \left[-\frac{2}{3}(\kappa^2-1)^{3/2}\right],
\end{equation*}
\begin{equation}
  Q\rightarrow 2\kappa \quad (C(z') \rightarrow 0). \label{T}
\end{equation}

Thus in this limit $A(z')$ is for finite $z'$ extremely large and negative, while $B(z')$ is extremely small and positive. The quantity $C(z')$ dies away.

Let us finally derive an expression for the field energy density $u(z)$, on the plateau $z>1$. We may start directly from the expression \cite{milton11}
\begin{equation}
u(z)=\frac{1}{2}\int \frac{d\zeta}{2\pi}\frac{d \bf k}{(2\pi)^2}\,e^{i\zeta \tau}\left(-\zeta^2+{\bf k}^2+\frac{\partial}{\partial z}\frac{\partial}{\partial z'}\right) g(z,z')\Big|_{z'\rightarrow z}.\, \label{U}
\end{equation}
where $\tau$ has been introduced as a regularization parameter. With
\begin{equation}
g(z,z')=\frac{1}{2\kappa}e^{-\kappa |z-z'|}-\left( \frac{1}{2\kappa}-\frac{1}{Q}\right) e^{-\kappa (z+z'-2)} \label{V}
\end{equation}
we get
\begin{equation}
u(z)=\frac{1}{2}\int \frac{d\zeta}{2\pi}\frac{d\bf k}{(2\pi)^2}\,e^{i\zeta \tau}\left[ -\frac{\zeta^2}{\kappa}-{\bf k}^2\left( \frac{1}{\kappa}-\frac{2}{Q}\right)e^{-2\kappa (z-1)}\right]. \label{W}
\end{equation}
Here we omit the first term, which is independent of the potential $V$, and which moreover is divergent when the regulator $\tau$ is put equal to zero. Introducing polar coordinates in the $\zeta \bf k$ volume, so that
\begin{equation}
\zeta =\kappa \cos \theta, \quad |{\bf k}|=\kappa \sin \theta, \label{X}
\end{equation}
we can write the volume element as $d\zeta {\bf k}=2\pi \kappa^2\sin \theta d\theta d\kappa$. With $\tau=0$ in the second term we can now perform the integration over $\theta$ from $0$ to $\pi$ to get, for $z\geq 1$,
\begin{equation}
u(z)=-\frac{2}{3}\frac{1}{(2\pi)^2}\int_0^\infty \kappa^4d\kappa \left(\frac{1}{\kappa}-\frac{2}{Q}\right) e^{-2\kappa (z-1)}. \label{Y}
\end{equation}
This expression is reasonably simple, and can be evaluated numerically when $Q$ is inserted from Eq.~(\ref{R}). The energy density goes to zero when $z\rightarrow \infty$, as expected. For finite $z$, the energy density is negative, similarly as in the case of a scalar field between two conducting plates. The sign is also in accordance  with the energy density obtained in Ref.~\cite{milton11}, Fig.~1, when the conformal parameter is put equal to zero. It may  be mentioned that for zero argument the Airy functions are known exactly,
\begin{equation}
{\rm Ai}(0)=\frac{1}{3^{2/3}\Gamma(2/3)}, \quad {\rm Bi}(0)=\frac{1}{3^{1/6}\Gamma(2/3)}. \label{Z}
\end{equation}
For large arguments $\kappa$, $(1/\kappa-2/Q) \rightarrow 0$, as  $Q \rightarrow 2\kappa$.

  \section{Final Remarks}
  The piecewise uniform string model, the main theme of this paper,  is a natural generalization
  of the conventional uniform string. The adaptability of the
  formalism to various regularization schemes - three of them considered above -
  should be emphasized. It is important to recognize that
  the assumption about relativistic
  invariance, shown in   Eq.~(\ref{1}), has to be satisfied in order for the formalism to work. If this
  assumption were removed, the regularization  would be difficult to
  handle; there would remain an ambiguity how to construct the counter term.

  Another point worth noticing is the close connection between
  the relativistic invariance property and the theory of an
  electromagnetic field propagating in an isorefractive
  medium meaning that  the refractive index is equal to one, or at least  a constant everywhere in  the material system
 \cite{brevik09,ellingsen09}. Again, if the isorefractive (or relativistic) condition were removed in the electromagnetic case, the regularization procedure would be rather difficult to deal with, as the contact term to be subtracted off would then depend on which of the media one chooses for this purpose.

 We recall that the quantization procedure in Sec.~5 was based upon two simplifying conditions. First, the tension ratio $x$ was taken to be small. This assumption implies that the eigenvalue spectrum for the composite string becomes simple: there are two branches, the first branch corresponding to $\omega_n(s)=(1+s)n$ and the second branch corresponding to $\omega_n(s^{-1})=(1+s^{-1})n$, with $n$ an integer. Our second assumption was that $s$ is an integer. We considered branch only, in detail.

One may ask: what is the Hagedorn temperature $T_c=1/\beta_c$ for the composite string? This temperature, introduced by Hagedorn in the context of strong interactions, is the temperature above which the free energy is ultraviolet divergent. Making use  of the Meinardus theorem \cite{meinardus54} this topic was discussed in Ref.~\cite{brevik03}. For the first branch we found
\begin{equation}
\beta_c=\frac{4\pi}{\sqrt{2(1+s)t(s)}}=4\sqrt{\frac{\pi}{2sT_{II}}}. \label{66}
\end{equation}
The important point here is that in the point mass limit, $s\rightarrow \infty$, one gets $\beta_c\rightarrow 0$, or $T_c \rightarrow \infty$. The Hagedorn temperature diverges in this limit.

Our second theme in this paper, the quantum field theory of the scalar field $\phi(z)$ in Sect. 7, is formally related to the string problem in the extreme case when the tension ratio $x\rightarrow 0$. This simple model deserves a study in its own right, as an example of the field theoretical models currently studied in the literature.

\newpage
{\bf Acknowledgement}

\bigskip

I thank Simen {\AA}. Ellingsen for computer help in connection with the discussion in Sect. 7.


\begin{thebibliography}{99}

\bibitem{brevik90}
Brevik, I. and Nielsen, H. B. 1990 {\it Phys. Rev. D} {\bf 41}, 1185.
\bibitem{li91}
Li, X., Shi, X. and Zhang, J. 1991 {\it Phys. Rev. D} {\bf 44}, 560.
\bibitem{brevik94}
Brevik, I. and Elizalde, E. 1994 {\it Phys. Rev. D} {\bf 49}, 5319.
\bibitem{elizalde95}
Elizalde, E. 1995 {\it Ten Physical Applications of Spectral Zeta Functions} (Berlin-Springer), Chapter 7.
\bibitem{brevik95}
Brevik, I. and Nielsen, H. B. 1995 {\it Phys. Rev. D} {\bf 51}, 1869.
\bibitem{brevik96}
Brevik, I., Nielsen, H. B. and Odintsov, S. D. 1996 {\it Phys. Rev. D} {\bf 53}, 3224.
\bibitem{bayin96}
Bayin, S. S., Krisch, J. P. and Ozcan, M. 1996  {\it J. Math. Phys.} {\bf 37}, 3662.
\bibitem{brevik97}
Brevik, I. and Sollie, R. 1997 {\it J. Math. Phys.} {\bf 38}, 2774.
\bibitem{berntsen97}
Berntsen, M. H., Brevik, I. and Odintsov, S. D. 1997 {\it Ann. Phys. NY} {\bf 257}, 84.
\bibitem{brevik98}
Brevik, I., Bytsenko, A. A. and Nielsen, H. B. 1998 {\it Class. Quant. Grav.} {\bf 15}, 3383.
\bibitem{brevik99}
Brevik, I., Elizalde, E., Sollie, R. and Aarseth, J. B. 1999 {\it J. Math. Phys.} {\bf 40}, 1127.
\bibitem{hadasz11}
Hadasz, L., Lambiase, G. and Nesterenko, V. V. 2000 {\it Phys. Rev. D} {\bf 62}, 025011.
\bibitem{brevik03}
Brevik, I., Bytsenko, A. A. and Pimentel, B. M. 2003 In: {\it Theoretical Physics 2002}, Part 2, p. 117. Eds.: T. F. George and H. F. Arnoldus (New York: Nova Sci. Publ.).
\bibitem{brevik11}
Brevik. I. 2011 In: {\it Cosmology, Quantum Vacuum and Zeta Functions}. In honor of Prof. Emilio Elizalde, Springer Proceedings in Physics 127, Eds. S. D. Odintsov, D. Saez-Gomez and S. Xambo-Descamps, p. 57 [arXiv:1007.1354].
\bibitem{lu98}
Lu, J. and  Huang, B. 1998 {\it Phys. Rev. D} {\bf 57}, 5280.
\bibitem{brevik09}
Brevik, I., Ellingsen, S. {\AA} and Milton, K. A. 2009 {\it Phys. Rev. E} {\bf 79}, 041120.
\bibitem{ellingsen09}
Ellingsen, S. {\AA}., Brevik, I. and Milton, K. A. 2009 {\it Phys. Rev. E} {\bf 80}, 021125.
\bibitem{feng10}
Feng, C. J. and Li, X. Z. 2010 {\it Phys. Lett. B} {\bf 691}, 167.
\bibitem{zhai11}
Zhai, X. H., Li, X. Z. and Feng, C. J. 2011 {\it Mod. Phys. Lett. A} {\bf 26}, 669.
\bibitem{zhai11a}
Zhai, X. H., Li, X. Z. and Feng, C. J. 2011 {\it Eur. Phys. J. C} {\bf 71}, 1654.
\bibitem{zhai11b}
Zhai, X. H., Li, X. Z. and Feng, C. J. 2011 arXiv:1107.4846 [hep-th].
\bibitem{fulling07}
Fulling, S. A., Kaplan, L. and  Wilson, J. H. 2007 {\it Phys. Rev. A} {\bf 76}, 012118.
\bibitem{harrison11}
Harrison, J. M. and Kirsten, K. 2011 {\it J. Phys. A: Math. Theor.} {\bf 44}, 235301.
\bibitem{vankampen68}
van Kampen, N. G., Nijboer, B. R. A. and Schram, K. 1968 {\it Phys. Lett. A} {\bf 26}, 307.
\bibitem{elizalde94}
Elizalde, E., Odintsov, S. D., Romeo, A., Bytsenko, A. A. and Zerbini, S. 1994 {\it Zeta Regularization Techniques with Applications} (Singapore: World Scientific).
\bibitem{brevik99a}
Brevik, I., Bytsenko, A. A. and  Goncalves, A. E. 1999 {\it Phys. Lett. B} {\bf
453}, 217.
\bibitem{hagedorn65}
Hagedorn, R, 1965 {\it Suppl. Il Nuovo Cimento} {\bf 3}, 147.
\bibitem{milton11}
K. A. Milton, Phys. Rev. D {\bf 84}, 065028 (2011),
\bibitem{bouas11}
J. D. Bouas, S. A. Fulling, F. D. Mera, K. Thapa, C. S. Trendafilova and J. Wagner, arXiv:1106.1162 [Proc. Symp. Pure Math., to be published].
\bibitem{NIST}
{\it NIST Handbook of Mathematical Functions}, edited by Frank W. J. Olver {\it et al.} (Cambridge University Press, 2010), Eqs.9.7.15 and 9.7.16.


\bibitem{meinardus54}
Meinardus, G. 1954 {\it Math. Z.} {\bf 59}, 338; {\bf 61}, 289.





\end{thebibliography}
\end{document}